 \documentstyle[12pt]{article}

 \input{epsf}

 \newcommand{\beq}{\begin{eqnarray}}
 \newcommand{\eeq}{\end{eqnarray}}

 \newcommand{\hal}[1]{{#1 \over 2}}
 \newcommand{\ket}{\rangle}
 \newcommand{\bra}{\langle}
 
 \newcommand{\bracket}[2]{\langle #1 | #2 \rangle}
 \newcommand{\pslash}{{p\hspace{-5pt}/}}
 
 \newcommand{\kslash}{{k\hspace{-6pt}/}}
 
 \newcommand{\gpNN}{g_{\pi N N}}
 \newcommand{\gpNR}{g_{\pi N N^*}}
 \newcommand{\gpRR}{g_{\pi N^* N^*}}
 \newcommand{\geNR}{g_{\eta N N^*}}

 \def\gsim{\displaystyle\mathop{>}_{\sim}}

 \def\Journal#1#2#3#4{{#1} {\bf #2}, #3 (#4)}

 \def\NPA{{\em Nucl. Phys.} A}

 \def\PRL{{\em Phys. Rev. Lett.}}
 \def\PRD{{\em Phys. Rev.} D}

\begin{document}

 \begin{center}
 {\large
 Probing chiral symmetry of nucleon by threshold
 $\eta \pi$ production\\ }
 \vspace*{0.5cm}

 {Daisuke Jido\\
 {\it Department of Physics, Kyoto University, Kyoto 606-8502 Japan}\\
 \vspace*{0.5cm}
 Makoto Oka\\
 {\it Department of Physics, Tokyo Institute of Technology,
 Meguro, Tokyo 152-8551 Japan}\\
 \vspace*{0.5cm}
 and\\
 \vspace*{0.5cm}
 Atsushi Hosaka\\
 {\it Numazu College of Technology, Numazu 410-8501, Japan} }

 \end{center}

 \vspace*{1cm}
 \abstract{
 Double meson production of the eta and pion at the threshold region
 is investigated in order to determine chiral property of the nucleon.
 The eta can be used as a probe for the
 negative parity
 nucleon $N^* \equiv N^*(1535)$ produced in the intermediate state.
 The coupling of the low energy pion in the final state
 is then used to extract the
 sign of the Yukawa coupling, $\gpRR$, which distinguishes the two
 realizations of chiral symmetry, either naive or mirror,
 for the nucleon.}

 \newpage

 \section{Introduction}

 Chiral symmetry with spontaneous breakdown is one of the
 important concepts in the dynamics of the strong
 interaction~\cite{Coleman}.
 In QCD chiral symmetry is realized as the symmetry in light flavors,
 $SU(N_{f})_{L} \times SU(N_{f})_{R}$
 for left- and right-handed quarks.
 Physical particles are then classified into appropriate
 representations of the chiral group which involve positive and
 negative parity states.
 For instance the pion and sigma can be assigned as members of the
 $(1/2, 1/2)$ representation, and $\rho$ and $a_{1}$ as the
 $(1,0) + (0,1)$ representation.
 These particles are then expected to be degenerate when chiral
 symmetry is restored.


 Chiral symmetry is realized at the hadronic level either in 
 the nonlinear or in the linear representations. The former is the 
 basis of  the chiral perturbation theory \cite{ChiP}, and is
 under well control to describe dynamics of mesons and
 baryons at low energy as the lagrangian
 is expanded in powers of small momenta. 
 The latter embodies the spontaneous
 breakdown of chiral symmetry. 
 Therefore, it is suited to the study of the 
 change in the vacuum toward the restoration of chiral 
 symmetry.  
 However, such a study, by regarding baryons explicitly 
 as representations of chiral symmetry, 
 has been 
 performed only by a limited number of
 authors~\cite{BWLee,DeTKun,CohJi}.
 In previous publications~\cite{JiOkHo,NJOH,JNOH},
 we have pointed out that there are
 two classes of linear representations for baryons
 when there are positive and negative parity baryons.
 We have called the one naive and the other mirror
 assignments~\cite{NJOH,mirror}.
 Physical implications of these two assignments are very
 different~\cite{NJOH,JNOH,KiJiOk}.
 In the naive assignment, the positive and negative parity nucleons
 belong to different chiral multiplets, and therefore,
 chiral symmetry does not relate them.
 In contrast,
 in the mirror assignment, they belong to the same chiral
 multiplet, where several nontrivial features emerge.
 For instance  nucleon masses can remain finite
 when chiral symmetry is restored, which is a feature that can not be
 possible in the naive assignment.


 The mirror assignment was first considered in literatures
 by Lee~\cite{BWLee}
 and later discussed by DeTar and Kunihiro~\cite{DeTKun}
 in some detail.
 Also an interesting suggestion was made recently by Jido, Kunihiro and
 Hatsuda~\cite{JiKuHa},
 where nucleon and
 delta resonances are put in the mirror
 chiral multiplet.
 There, observed masses and decay strengths are
 remarkably consistent with
 predictions of a simple linear sigma model.
 Nevertheless,
 so far, we can not clearly
 distinguish these chiral assignments in physical
 nucleons, since observations for key quantities are rather
 difficult.

 One thing which differs in the two chiral assignments
 is the relative sign of the axial charges
 ($g_{A}$ and $g_{A}^*$), or equivalently the pion
 coupling ($\gpNN$ and $\gpRR$)
 for the positive and negative parity nucleons.
 The difference in the sign
 may be observed through interference effects.
 The purpose of the present paper is
 to propose one of such processes,
 which we hope will be performed in future experiments.

 If we assume that the relevant negative parity nucleon is
 $N(1535)$, the first excited state of negative parity, it can
 be probed by the production of the eta meson.
 Therefore, we are lead naturally to the study of the $\pi \eta$
 production especially at the threshold region.
 As an illustration, we investigate in this paper
 a pion induced process, $\pi^- + p \to \pi^- +  \eta + p$.
 This reaction would be relatively in easy access by experiments,
 since the charged particles ($\pi^-$ and $p$)
 are present and therefore the $\eta$
 in the final state can be observed by invariant mass
 analysis of the $\pi^- p$ system.

 The contents of this paper are the followings.
 In the next section we briefly review the chiral symmetry for the
 nucleon.
 The physical consequences of the naive and mirror assignments are
 summarized.
 In section 3, we formulate the two meson productions of the pion and
 eta at the threshold region.
 In section 4, we present several kinds of cross sections and discuss
 how the two assignments differ in various observed quantities.
 Conclusion is given in section 5.

 \section{Chiral symmetry of the nucleon}

 Let us start with a brief review on chiral representations of the
 nucleon~\cite{JNOH}.
 %
 To be specific, we consider the chiral group with two flavors,
 $SU(2)_{L} \times SU(2)_{R}$.
 The nucleon field $N$ is decomposed into the left and right handed
 components, $N = N_{l} + N_{r}$.
 Assuming linear representations of the chiral group for the nucleon,
 $N_{l}$ and $N_{r}$ form the two fundamental representations of the
 chiral group, $(1/2, 0)$ and $(0,1/2)$, respectively, where the first
 (second) number in the parentheses refers to the representation
 of $SU(2)_L$ ($SU(2)_R$).
 Chiral transformations are isospin
 transformations for the left and right handed nucleons as
 $N_{l} \to g_{L} N_{l}$ and $N_{r} \to g_{R} N_{r}$, where
 $g_{L} \in SU(2)_{L}$ and $g_{R} \in SU(2)_{R}$.
 Hence, the transformation on the left- or right-handed nucleon is
 called the left ($L$) or right ($R$) chiral transformation.
 For a massive nucleon, the mass term must couple
 the left and right handed components (the Dirac mass
 term) as $m\bar NN = m(\bar N_{l} N_{r} + \bar N_{r}N_{l})$.
 Hence chiral symmetry must be broken.

 When there are two different nucleon fields the situation can change.
 Let us denote the two nucleons as $N^{+}$ and $N^{-}$.
 Then we can introduce two
 different ways of chiral assignments for the left and right handed
 components, as shown in Table~1.
 Here the superscripts $+$ and $-$ imply the parity of the two nucleons.
 In the naive assignment, the left- and right-handed components of
 the two nucleons, $N^{+}$ and $N^{-}$, behave in the same way,
 while in the mirror assignment, the roles of the
 left- and right-handed components of the second nucleon $N^-$ are
 interchanged, just as in the mirror world
 where the left and right hand sides are interchanged.  
 The reason that such an assignment is possible is that the chiral
 symmetry is an internal symmetry and when there are two (or generally
 more) different nucleon fields, the left and right handed
 components of different nucleons
 do not necessarily behave in the same way.
 In the mirror assignment, due to the opposite transformation property,
 the chiral invariant mass term is allowed,
 $m(\bar N^{+} \gamma_{5}N^{-} - \bar N^{-} \gamma_{5}N^{+})
 = m(\bar N^{+}_{l} N^{-}_{r} + \bar N^{-}_{r} N^{+}_{l}
 - \bar N^{-}_{l} N^{+}_{r} - \bar N^{+}_{r} N^{-}_{l})$.


 \begin{table}[tbp]
         \centering
         \caption{\small Two chiral assignments for two nucleons.}
     \vspace*{0.5cm}
         \begin{tabular}{ c c c c | c c c c }
                 \hline
                  \multicolumn{4}{c |}{Naive} &
                  \multicolumn{4}{c}{Mirror}   \\
                 \hline
                 $N_{l}^+$ & $N_{l}^-$ & $N_{r}^+$ & $N_{r}^-$ &
                 $N_{l}^+$ & $N_{r}^-$ & $N_{r}^+$ & $N_{l}^-$\\
                 \multicolumn{2}{c}{(1/2,0)} &
                 \multicolumn{2}{c |}{(0,1/2)} &
                 \multicolumn{2}{c}{(1/2,0)} &
                 \multicolumn{2}{c}{(0,1/2)} \\
                 \hline
         \end{tabular}
         \label{chitrans}
 \end{table}

 In our previous publication, based on these classifications,
 we have investigated several physical implications of the two
 assignments using linear sigma models~\cite{JNOH,KiJiOk}.
 Main results are summarized in Table~2.
 The masses of the nucleons in the Wigner phase should be absent
 in the naive case, while they can be finite in the mirror case. It would be
 interesting to study physics near the chiral phase transition. The $\pi NN^*$
 coupling in the chiral limit should vanish at the tree level in the naive
 assignment, which is qualitatively supported by the small observed
 value, $g_{\pi NN^*} \sim 1$.
 The off-diagonal
 axial charge $g_A^{N N^*}$ at finite temperature and/or density should be
 suppressed in the naive case, while enhanced in the mirror case,
 as compared to that at the normal vacuum. This is because $g_A^{N^* N^*}$ should
 approach one in the naive case, while zero in the mirror case, as the
 broken chiral symmetry tends to be restored.

 In the present work, among various properties listed there, we would like
 to discuss a possibility to measure
 the sign of the axial charge which is perhaps the most clearcut signal
 to distinguish the two chiral assignments.
 The sign of the axial charges, when combined with the
 Goldberger-Treimann relation, can be turned into the sign of the
 Yukawa couplings.
 Then the difference in the sign of the couplings $g_{\pi N^{+} N^{+}}$ and
 $g_{\pi N^{-} N^{-}}$ could be
 observed in a suitable hadronic processes which involve the pion.

 In the following discussion we assume that candidates of the two
 nucleons are the ground state nucleon $N(938)$ and the first negative
 parity excited state $N(1535)$.
 It is known that $N(1535)$ couples strongly to the eta meson, which
 can be used as an indication that $N(1535)$ is produced in relevant
 reactions.
 In what follows we often denote $N(939)$ simply by $N$, and $N(1535)$
 by $N^*$.

 \begin{table}[tbp]
         \centering
         \footnotesize
 \caption{\label{sum} Comparison between the naive and
 mirror assignments.}
 \vspace*{0.5cm}
 \begin{tabular}{l |c c}
 \hline
       & Naive assignment & Mirror assignment\\
 \hline
 Chiral multiplet & $(N^{+}, \gamma_{5} N^{+})$  and
                    $(N^{-}, \gamma_{5} N^{-})$  &
        $(N^{+}, \gamma_{5} N^{+}, N^{-}, \gamma_{5}N^{-})$\\
        & $N^+$ and $N^-$ are independent &  \\
 Mass in the Wigner phase& 0 & $m_0$ (finite) \\
 $\pi NN^* $ coupling & 0 & $ (a + b)/\cosh \delta$ \\
 Relative sign of $g_{A}^{NN}$ and $g_{A}^{N^* N^*} $
 & Positive & Negative  \\
 $g_{\pi NN} (\rho, T)$ & Decrease & Decrease \\
 $\gpNR (\rho, T)$ & Increase & Decrease \\
 \hline
 \end{tabular}

 \end{table}

 \section{Formulation}

 We investigate an eta pi production process $\pi^- + p \rightarrow \pi^- +
 \eta + p$, in order to extract the relative sign of $\gpRR$ to $\gpNN$.
 As indicated in introduction, our basic idea is to see
 interference effects
 due to sign difference between the two coupling constants,
 $\gpNN$ and $\gpRR$.
 Let us see the diagrams (1) and (2) in Figs.1. The $\pi NN$ coupling
 is in the diagram (1), while the $\pi N^* N^*$ coupling in the diagram (2).
 The relative sign of the couplings results in either
 constructive or destructive sum of these two diagram.

\begin{figure}[tbp]
    \vspace*{1cm}
    \centering
    \footnotesize
    \epsfxsize = 11cm
    \epsfbox{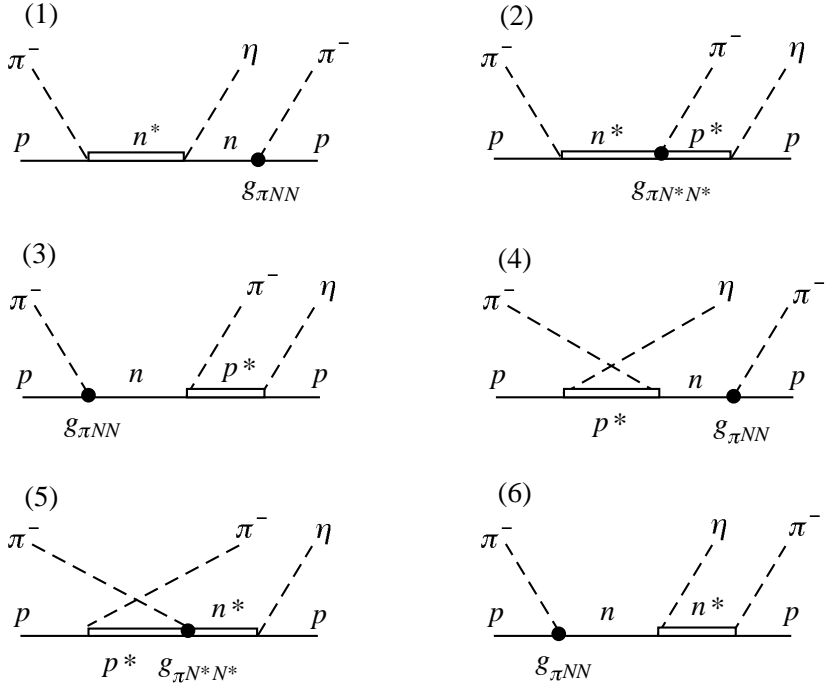}
  \begin{minipage}{12cm}
    \centering
    \caption{ \small
    Six pole dominant diagrams for $\eta \pi$ production.
    \label{sixdiag}}
 \end{minipage}

 \end{figure}

 Our assumptions are the followings
 \begin{enumerate}
         \item  Resonance ($N(1535)$)  pole dominance.
         This is considered to be good
         particularly for the $\eta$ production process at the threshold
         region.

         \item  The $\eta$ meson couples only through $N(1535)$.
         It is known that the $\eta$ meson couples only weakly with the
         nucleon and other resonances.

     \item  Pions are inserted
     in all possible ways to the diagrams.
 \end{enumerate}
 Under these assumptions, we can write six diagrams as shown
 in Figs.~\ref{sixdiag}.

 For meson nucleon interactions,
 we take the interaction lagrangians:
 \beq
 L_{\pi NN}
 &=& \gpNN \bar N i \gamma_{5} \vec \tau \cdot \vec \pi N \, ,
 \nonumber \\
 L_{\eta NN^*}
 &=& \geNR ( \bar N \eta N^*  + \bar N^* \eta N )  \, ,
 \nonumber \\
 L_{\pi NN^*}
 &=& \gpNR ( \bar N \tau \cdot \pi  N^*
            + \bar N^* \tau \cdot \pi N )  \, ,
 \nonumber \\
 L_{\pi N^*N^*}
 &=& \gpRR ( \bar N^* i \gamma_{5} \tau \cdot \pi  N^* )  \, .
 \label{Lints}
 \eeq
 Here we have adopted pseudo-scalar couplings which arise naturally in
 linear sigma models.
 We use these interactions both for the naive and mirror cases with
 empirical coupling constants for $\gpNN \sim 10$, $\gpNR \sim 0.7$
 and $\geNR \sim 2$.
 The second and third are determined from the partial decay widths,
 $\Gamma_{N^*(1535) \to \pi N} \approx  \Gamma_{N^*(1535) \to \eta N}
 \sim 70$ MeV~\cite{PDG}, although large
 uncertainties for the width have been reported~\cite{ManSal,VrDtLe}.
 The unknown parameter is the $\gpRR$ coupling.
 One can estimate it by using the theoretical value of the
 axial charge $g_{A}^{*}$ and the Goldberger-Treimann relation for $N^*$.
 When $g_{A}^{*} = \pm 1$ for the naive and mirror assignments,
 we find
 $
 \gpRR = g_{A}^{*} m_{N^*}/{f_{\pi}} \sim \pm 15 .
 $
 Here, just for simplicity, we use the same absolute value as
 $g_{\pi NN}$.
 The coupling values
 used in our computations are summarized in
 Table~\ref{parameters}.

 \begin{table}[tbp]
         \centering
 \caption{\label{parameters} \small Parameters used in our calculation. }
 \vspace*{0.5cm}
 \begin{tabular}{ c c c c c c c }
 \hline
 $m_{N}$ & $m_{N^*}$ & $\Gamma_{N^*}$ & $\gpNN$
    & $\gpNR$ & $\geNR$ & $\gpRR$  \\
 \hline
 938 & 1535  & 140            & 13
    & 0.7     &  2.0    &  13 (naive) \\
 (MeV) & (MeV) & (MeV) &
    &         &         &  --13 (mirror) \\
 \hline
 \end{tabular}
 \end{table}

\begin{figure}[tbp]
    \vspace*{1cm}
    \centering
    \footnotesize
    \epsfxsize = 13cm
    \epsfbox{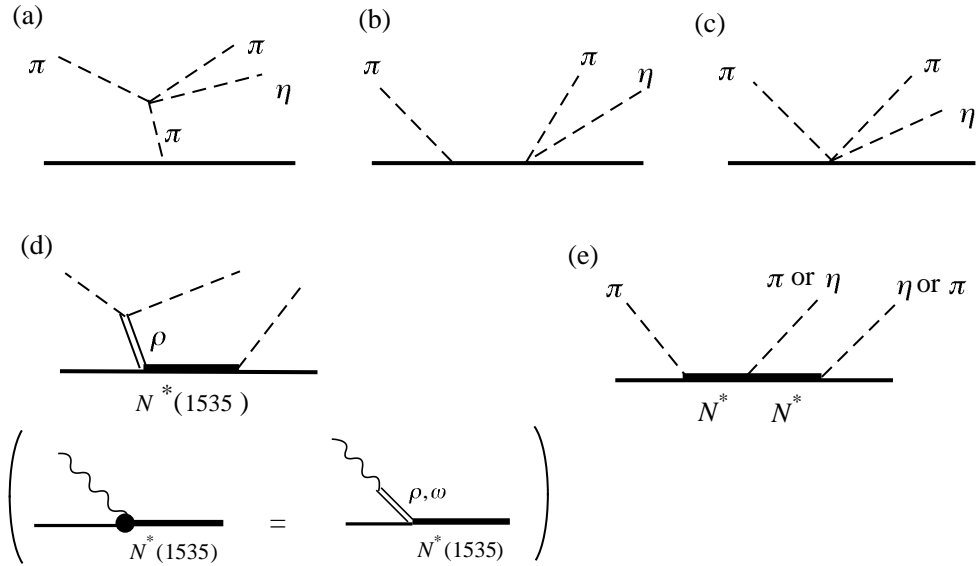}
  \begin{minipage}{12cm}
    \centering
    \caption{ \small
    Various contributions to $\pi N \to \eta \pi N$.
    \label{bkground}}
 \end{minipage}

 \end{figure}

 Besides the resonance pole contributions as shown in
 Fig.~\ref{sixdiag}, there are several other possible terms.
 We ignore all of them from the following reasons.
 \begin{itemize}
         \item  {\bf Background:}\\
     We can consider three diagrams as shown in Figs.~\ref{bkground}
     (a-c).
     All of them are not allowed due to G-parity.
         For (a), the vertex $\pi \pi \pi \eta$ is G-parity forbidden, since
         $G(\pi) = -1$ and $G(\eta) = 1$.
         For (b) and (c), to estimate the diagrams, we first consider
         the lowest order of the chiral expansion in the Lagrangian~\cite{DoGoHo92}.
         For two-meson nucleon vertices in (b), the two mesons are
         correlated as vector mesons (such as $\rho$) which have G-parity
         plus.
         Therefore, the G-parity minus combination $\pi \eta$ is not
         allowed.
         Similarly, three meson vertices in (c) have axial vector
         correlation with negative G-parity,
         and hence the positive G-parity combination $\pi \pi \eta$ is
         not possible.
         These selection rules are explicitly satisfied in actual chiral
         lagrangians.

         \item  {\bf $\rho$ meson:}\\
         We have computed the diagram in Fig.~\ref{bkground} (d) explicitly.
         The rho meson coupling to $N$ and $N^*$ is extracted from the helicity
         amplitudes $A_{1/2} \sim 0.08 \; GeV^{-1/2}$~\cite{PDG}
         using the vector meson dominance as shown in Fig.~\ref{bkground} (d).
         It turns out that the contribution to the cross section
         is negligibly small as compared
         to the resonance pole terms in Fig.~\ref{sixdiag}
         by about factor $10^{-3}$.

         \item  {\bf Off diagonal couplings:}\\
         Finally, one would expect contributions where two resonances appear
         in intermediate states as shown in Fig.~\ref{bkground} (e).
         Again we can ignore these diagrams safely,
         since there is no strong indication
         that any resonances couples to $N^*(1535)$ by emitting a
         pion~\cite{PDG}.
         In particular, the delta resonance which could be excited strongly
         by the incident pion does not couple to $N^*(1535)$, since the
         observed branching ratio of $N^*(1535) \to \Delta \pi$ is less than
         10 \%.
 \end{itemize}

\begin{figure}[tbp]
    \vspace*{1cm}
    \centering
    \footnotesize
    \epsfxsize = 8cm
    \epsfbox{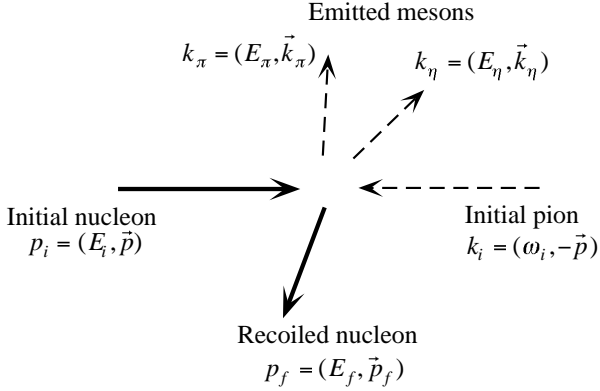}
  \begin{minipage}{12cm}
    \centering
    \caption{ \small
    Definition of momentum variables. \label{momenta}}
 \end{minipage}

 \end{figure}

 Now the $S$-matrix is computed using the reduction formula
 \beq
 S_{fi} &=&
 _{out}\bra p_{f}, k_{\pi}, k_{\eta} | k_{i}, p_{i}\ket_{in}
 \nonumber \\
 &=&
 disc.  + (iZ_{\pi}^{-1/2})^2 (iZ_{\eta}^{-1/2})
 \int d^4x d^4y d^4z \,
 e^{ik_{\pi}x + ik_{\eta}y - ik_{i} z} \nonumber \\
 & & (\Box_{x} + m_{\pi}^2)  (\Box_{y} + m_{\eta}^2)
 (\Box_{z} + m_{\pi}^2)
 _{out}\bra p_{f}|T(\pi^i(x) \eta(y) \pi^j(z)) | p_{i}\ket_{in} \, .
 \label{smatrix}
 \eeq
 The momentum variables are defined as in Fig.~\ref{momenta}.

 In the perturbation theory, the matrix element can be computed as
 \beq
 & & _{out}\bra p_{f}|T(\pi^i(x) \eta(y) \pi^j(z)) | p_{i}\ket_{in}
 \nonumber \\
 & & \; = \;
 i^3 \bra p_{f}|T ( \pi^i(x) \eta(y) \pi^j(z)
 \int d^4x_{1} d^4x_{2} d^4x_{3}
 L_{\eta NN^*}(x_{1}) L_{\pi NN^*}(x_{2}) \nonumber \\
 & & \hspace*{2cm}
 \times (L_{\pi NN}(x_{3}) + L_{\pi N^*N^*}(x_{3})) )
 | p_{i}\ket_{in} \, .
 \label{ptbn}
 \eeq
 After taking the Wick contraction in (\ref{ptbn}) which is then
 inserted in (\ref{smatrix}), we find an ordinary expression for
 amplitudes in momentum space.
 For instance the amplitudes for Fig.~\ref{sixdiag}
 (1) and (2) are given by
 \begin{eqnarray}
     {T}{(1)} &=& \bar{u}(p_{f}) { (i\sqrt{2}g_{\pi NN}i\gamma_{5})i
     (i g_{\eta NN^{*}}) i( i \sqrt{2} g_{\pi NN^{*}})\over
     (\pslash_{f} + \kslash_{\pi} - m_{N})( \pslash_{i}+\kslash_{i} -
     m_{N^{*}} + \hal{i} \Gamma )}u(p_{i}) \, , \\
     {T}{(2)} &=& \bar{u}(p_{f}) {(i g_{\eta NN^{*}} ) i(i
     \sqrt{2} g_{\pi N^{*}N^{*}} i \gamma_{5})i (i \sqrt{2} g_{\pi
     NN^{*}}) \over (\pslash_{f} + \kslash_{\eta} - m_{N^{*}} + \hal{i}
     \Gamma) (\pslash_{i} + \kslash_{i} - m_{N^{*}} +\hal{i} \Gamma)
     }u(p_{i}) \, ,
 \end{eqnarray}
 where $u$'s are the Dirac spinors for the nucleon.

 Using these $\cal{T}$ matrices, we calculate cross section as
 \beq
 d\sigma &=& \frac{2m_{N}}
 {4 \sqrt{(p_{i} \cdot k_{i})^2 - m_{N}^2 m_{\pi}^2}} \frac{1}{2}
 \sum_{spin} |{\cal T}_{fi} |^2 d \Phi \, ,
 \label{crosssection}
 \eeq
 where the phase space of the three body final state is given by
 \begin{equation}
     d\Phi =
     (2\pi)^{4}  \delta(p_{i}+k_{i}-p_{f}-k_{\pi}-k_{\eta})
     {d^{3}k_{\pi} \over (2\pi)^{3} 2 E_{\pi}}
     {d^{3}k_{\eta} \over (2\pi)^{3} 2 E_{\eta}}
     {m_{N} d^{3}p_{f} \over (2\pi)^{3} E_{f}} \, .
 \end{equation}
 Here the convention for the normalization is
 \begin{eqnarray}
     \bar{u}^{(\alpha)}(p) u^{(\beta)}(p) & = & \delta^{\alpha\beta}
     \, ,  \\
     \bracket{p}{p^{\prime}} &=& {E \over m} (2\pi)^{3} \delta^{3}
     (\vec{p}-\vec{p}^{\prime}) \, .
 \end{eqnarray}
 In the center of mass frame, the phase space integral reduces to
 \begin{equation}
     d\Phi = \frac{m_{N}}{4(2\pi)^{5}} dE_{\pi} dE_{f}
     d\alpha d(\cos\beta) d\gamma \, .
 \end{equation}
 In the center of mass frame, the momenta of the emitted particles,
 $\vec{p}_{f}$ $\vec{k}_{\pi}$ and $\vec{k}_{\eta}$, lie in a plane.
 If the energies of the proton and the pion in the final state, $E_{f}$,
 $E_{\pi}$ are fixed, then the relative angles between either two of
 $\vec{p}_{f}$,
 $\vec{k}_{\pi}$ and $\vec{k}_{\eta}$ can be determined.
 Therefore, the orientations of the three momenta are
 specified by the three
 Euler angles $\alpha$, $\beta$ and $\gamma$.

 We have computed the integral over the three body phase space in
 the Monte Carlo method.
 The number of configurations is taken more than 30,000, depending
 on the kinds of cross sections.
 The total cross section is computed in a schematic way as
 \begin{eqnarray}
         \sigma & = & {\rm (K.F.)} \int |{\cal T(\xi)}|^{2}  d\Phi
         \nonumber \\
         & \to &  {\rm (K.F.)}{1 \over N} \sum_{i=1}^{N}
         |{\cal T}(\xi_{i})|^{2} V\, ,
 \end{eqnarray}
 where $\xi=(E_{f}, E_{\pi}, \alpha, \cos\beta, \gamma)$.
 The volume of the phase space is given by the integral
 \begin{equation}
 \label{volint}
         V = {m_{N} \over 4 (2 \pi)^{5}} 4 \pi^{2} \int_{E_{min}}^{E_{max}}
         {2 \sqrt{(E^{*2}_{\eta}- m_{\eta}^{2})(E^{*2}_{\pi}-m_{\pi}^{2})}
         \over E_{cm}} dE_{\pi} \, ,
 \end{equation}
 where $E^{*}_{\eta}$ and $E^{*}_{\pi}$ are the energies of the emitted $\eta$
 and $\pi$ in the rest frame of the emitted nucleon and eta.
 They are
 \begin{eqnarray}
         E^{*}_{\eta} & = & { E_{cm}^{2}-m_{N}^{2} + m_{\eta}^{2} +
         m_{\pi}^{2} - 2 E_{cm} E_{\pi} \over 2\sqrt{E_{cm}^{2} +
         m_{\pi}^{2} - 2E_{cm} E_{\pi}}} \, , \\
         E^{*}_{\pi} & = & {E_{cm} E_{\pi} - m_{\pi}^{2} \over
         \sqrt{E_{cm}^{2} + m_{\pi}^{2} - 2 E_{cm} E_{\pi}}} \, .
 \end{eqnarray}
 The lower and upper bounds of the integral (\ref{volint})
 are given by
 \begin{eqnarray}
         E_{min} & = & m_{\pi} \, , \\
         E_{max} & = & { E_{cm}^{2} + m_{N}^{2} - (m_{N} + m_{\eta})^{2}
         \over 2 E_{cm} }\, .
 \end{eqnarray}

 To compute differential cross section $d\sigma(\zeta)$ where
$\zeta$ is a representative of the variable we need,
for instance the angle of the emitted pion in the center of mass frame and
the momentum of the emitted pion in the laboratory frame, 
we put a delta function of finite range $\Delta \zeta$ in the
 integrand of the total cross section.
 \begin{eqnarray}
         d\sigma (\zeta)& = &{\rm (K.F.)} \int |{\cal T}(\xi)|^2
         \delta(\zeta^\prime (\xi) - \zeta) d \Phi \nonumber \\
         & \to &  {\rm (K.F.)}{1 \over N} \sum_{i=1}^{N}
         |{\cal T}(\xi_{i})|^{2} { \sqrt{\pi} \over \Delta\zeta^{2}}
         e^{-{(\zeta^\prime(\xi_i)-\zeta)^2 \over \Delta\zeta^2}}
         V\, .
 \end{eqnarray}
 Note that the phase space is represented in the center of mass frame.
 The $\zeta^\prime$ stands for the translation of the variables $\xi$.
 For example, $\zeta^\prime$ is a boost transformation from the CM frame to 
the laboratory frame to calculate the differential cross section in the 
 laboratory frame.

%
 \section{Results and discussions}

\begin{figure}[tbp]
    \vspace*{1cm}
    \centering
    \footnotesize
    \epsfxsize = 8cm
    \epsfbox{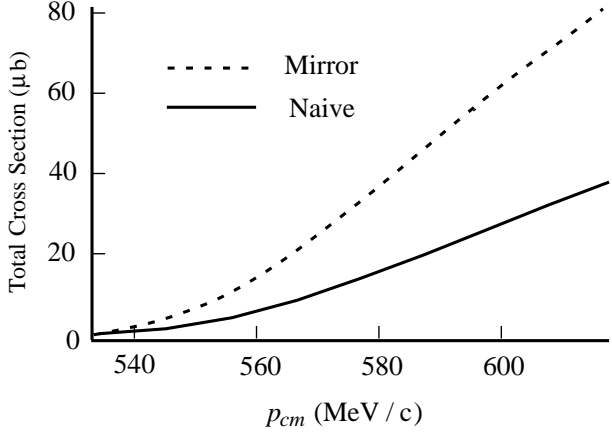}
  \begin{minipage}{12cm}
    \centering
    \caption{ \small
    Total cross sections of
    $\pi^- p \to \eta \pi^- p$ for the naive and mirror models as
    functions of the initial pion momentum
    $P_{c.m.}$ in the center of mass frame.
    \label{total}}
 \end{minipage}

 \end{figure}

 The total cross sections is shown in Fig.~\ref{total}
 as functions of
 the energy of the initial pion for the naive and mirror cases.
 The difference between the two is due to the sign
 of the $\gpRR$ coupling.
 The cross sections increase, as the initial pion energy and
 correspondingly the phase space of the final three body state
 increase.
 For $P_{cm} \gsim P_{cm}^{\rm thresold} + 50$ MeV/c
 ($P_{cm}^{\rm threshold} = 528$ MeV/c), the
 cross sections reach more than ten micro barn, which
 will be experimentally accessible.
 In the whole energy region as shown in Fig.~\ref{total}
 the cross section is larger in the mirror model,
 which is about twice as that in the naive model.

\begin{figure}[tbp]
    \vspace*{1cm}
    \centering
    \footnotesize
     \epsfxsize = 8cm
    \epsfbox{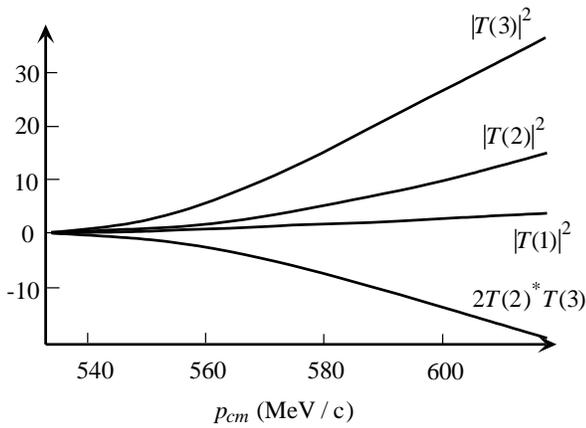}
  \begin{minipage}{12cm}
    \centering
    \caption{ \small
    Separate contributions from various terms of Fig.~\ref{sixdiag}.
    The amplitude of the cross term $2T(2)^{*} T(3)$ corresponds to the
    naive assignment.  It changes the sign for the mirror assignment.
    \label{various}}
 \end{minipage}

 \end{figure}

 Among various terms shown in Fig.~\ref{sixdiag}, major contributions
 are from the diagrams (2) and (3).
 In Fig.~\ref{various}, we show relative strengths of
 $|T(2)|^2$, $|T(3)|^2$ and their interference $2 T(2)^{*} T(3)$ in the
 naive assignment.
 The difference between the naive and mirror assignments in the total
 cross section is from the sign difference of this term.
 The third major contribution is from $T(1)$ which is also shown
 there.
 Other terms are negligible.

 The term $T(3)$ gives a large contribution
 to the cross section as compared
 with $T(1)$ and $T(2)$, although we expect naively that
 the terms $T(1)$ and $T(2)$ are the dominant contribution with
 considering of their energy denominators. This is so because of
 the p-wave nature of the $\pi NN$ coupling
 at the initial state of the diagram 3. The $\pi NN$ and $\pi N^* N^*$
 are reduced to the p-wave couplings in the non-relativistic limit.
 Near the threshold the emitted pion has a comparably small momentum.
 In the diagram 1 and 2 the p-wave coupling is attached on the emitted
 pion. Therefore it gives a suppression to their contributions, while
 in the diagram 3 the p-wave coupling is on the initial pion, which
 has a larger momentum than the emitted pion.


 The total cross section alone is not sufficient to distinguish
 the difference between the naive and mirror assignments, because
 it is hard to determine the absolute values of the cross section
 in experiments and there are the unknown parameter $\gpRR$. Therefore
 some qualitative differences are desired. Then let us discuss
 differential cross sections.

\begin{figure}[tbp]
    \vspace*{1cm}
    \centering
    \footnotesize
    \epsfxsize = 15cm
    \epsfbox{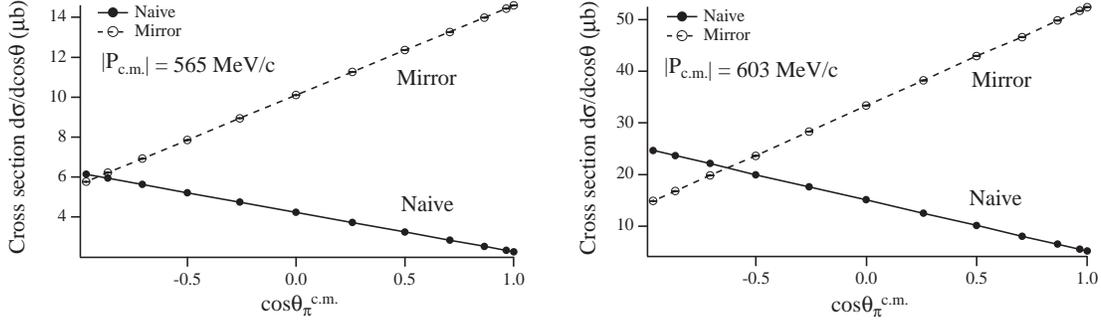}
  \begin{minipage}{12cm}
    \centering
    \caption{ \small
    Angular distributions of the
    $\pi^-$ in the final state. \label{ang}}
 \end{minipage}

 \end{figure}

\begin{figure}[tbp]
    \vspace*{1cm}
    \centering
    \footnotesize
    \epsfxsize = 15cm
    \epsfbox{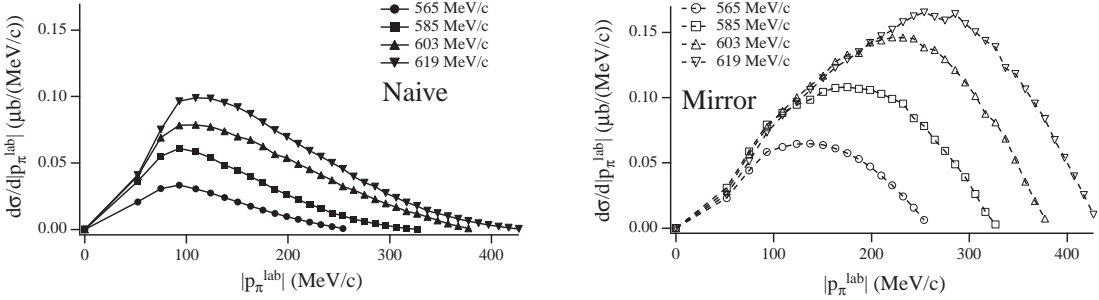}
  \begin{minipage}{14cm}
    \centering
    \caption{ \small
    Momentum distributions of the
    $\pi^-$ in the final state. \label{piene}}
 \end{minipage}

 \end{figure}

 \begin{itemize}
         \item  {\bf Angular distributions:}
         In Fig.~\ref{ang}, we show angular distributions of the final
         state  pion in the center of mass frame.
         This shows the clear difference between the two models. This comes
         from the p-wave nature of the $\pi N^*N^*$ coupling in the diagram 2.
         As it is mentioned before, the main contributions are given by the
         diagram 2 and 3, and difference by the sign of $\gpRR$ appears in the
         sign of the interference of $T(2)$ and $T(3)$. The first
         nucleon in the intermediate state is rest in the center of
         mass frame.
         Due to the structure of the Yukawa vertex, the emitted pion
         in the diagram 2 is in
         p-state, which gives monotonic increase or decrease in the
         differential cross sections  as functions of
         $\cos \theta_{\pi}^{c.m.}$. On the other hand, the emitted pion in
         the diagram 3 is in s-wave, which has no angular distributions.
         Therefore the cross term of $T(2)$ and $T(3)$ behaves 
         linearly in $\cos \theta_{\pi}^{c.m.}$, and
         the apparently different behaviors in the
         $\cos \theta_{\pi}^{c.m.}$
         dependence is due to the difference in the sign
         of the coupling $\gpRR$.
         This angular dependence would be one of the cleanest observables to
         distinguish the naive or mirror assignments.

         \item  {\bf Momentum distributions:}
         Another example which is useful
         is the momentum (energy) distribution of one of the final state
         particles.
         We plot the momentum distribution
         of the emitted pion in the laboratory frame in
         Fig.~\ref{piene} for several incident energies.
         What differs in
         the two chiral assignments is the position of the
         peak in the cross sections.
         In the naive case, it does not depend on
         the incident energy, while in the mirror case, it shifts to higher
         momentum region as the incident energy is increased.
 \end{itemize}


 \section{Conclusions}

 We have proposed the two meson production reaction
 $\pi^- p \to \pi^- \eta p$ to probe the chiral symmetry for the
 nucleon.
 Since chiral symmetry plays essential roles in hadron physics,
 its manifestation in the baryon sector as well as in
 the meson sector is
 very important to understand the nature of the QCD vacuum.
 Our aim is to extract the relative sign of the two strong
 coupling constants, $\gpNN$ and $\gpRR$ through their interference,
 since it is a signal to distinguish
 the chiral representations of the nucleons.

 We have investigated various cross sections for
 $\pi^- p \to \pi^- p \eta$.
 Having in mind possible experimental setups
 we have shown total cross sections, angular distributions of the pion
 and energy distributions of the pion.
 In fact, we have also computed differential cross sections as
 functions of the final proton angles and momenta.
 However, we did not see very clear distinction between the naive and
 mirror chiral assignments.
 The three examples we have presented here, the total cross section,
 angular and momentum distributions of the emitted pion, are the
 processes which show the most visible difference in the two assignments.

 In addition to the reaction we have considered in the present paper,
 there are similar ones such as
 $\pi^+ p \to \pi^+ \eta p$,
 $\gamma p \to \pi^0 \eta p$, and so on.
 In the former, as the total isospin is $I = 3/2$,
 the $\Delta(1232)$ channel would be dominant and therefore,
 interference effects among various terms may not be expected.
 In the latter, the final state contains two neutral particles
 ($\eta$ and $\pi^0$) involved, which
 makes experimental setup difficult.

 Theoretical predictions have been made under the
 assumption of resonance dominance of $N^*(1535)$.
 The unique feature of the resonance which strongly couples
 to $\eta$ makes the present theoretical analysis rather simple, since
 $\eta$ can filter only a limited number of diagrams.  
 Perhaps,
 the physical as well as practical (experimental) conditions select
 almost uniquely the present reaction as the most
 convenient one.

 We have then shown that various cross sections differ
 significantly depending on
 whether the nucleons belong to the naive or mirror chiral
 assignments.
 Not only a single but also several observations for different
 quantities will be useful to obtain information on the chiral symmetry
 for the nucleon.

 \section*{Acknowledgments}

 We acknowledge H. Kim and M. Iwasaki for discussions.
 This work is supported in part by the Grant-in-Aid for scientific
 research (C)(2) 11640261.

 \end{document}